**Biomaterial design inspired by membraneless organelles**

Jianhui Liu, Fariza Zhorabek and Ying Chau*

Department of Chemical and Biological Engineering

The Hong Kong University of Science and Technology

Clear Water Bay, Kowloon, Hong Kong

**Abstract**

Compartmentalization is ubiquitous in the broad cellular context, especially in the formation of membraneless organelles (MOs). Membraneless organelles (MOs) are phase-separated liquid compartments that provide spatiotemporal control of biomolecules and metabolic activities inside a cell. While MOs exhibit intriguing properties such as efficient compositional regulation, thermodynamic metastability, environmental sensitivity and reversibility, their formation is driven by weak non-covalent interactions derived from simple motifs of intrinsic disordered proteins (IDPs). Understanding the natural design of IDPs and the liquid-liquid phase separation behavior will not only reveal insights about the contributions of MOs to cellular physiology and disease pathology, but also provides inspirations for the *de novo* design of dynamic biomolecules depots, self-regulated biochemical reactors, and stimuli-responsive systems. In this article, the sequence and structural features of IDPS that contribute to the organization of MOs are reviewed. Artificial MOs formed following these principles, including self-assembling peptides, synthetic IDPs, polyelectrolytes and peptide-polymer hybrids are described. Finally, we illustrate the applications and discuss the potential of the MO-inspired biomaterials, with examples spanning biochemical reactors, synthetic biology, drug discovery and drug delivery.

**Table of Contents**

1. **Introduction**

Living cell metabolism demands compartmentalization for the precise spatiotemporal control of a myriad of biochemical processes[1]. Well-defined compartments host collections of specifically organized biomolecules, engendering a hierarchy of biological materials and the coordination of their biochemical activities. Compartmentalization is ubiquitous in the broad cellular context, both intra- and extracellularly, including membrane-bound organelles, extracellular vesicles (EVs) and membraneless organelles (MOs). Intracellularly, other than the "canonical organelles" bound by lipid bilayer membranes, MOs are where concentrating of bioactive contents takes place by their partition in separated liquid phases [2–7]. MOs are found in the cytoplasm (e.g. P granules and stress granules) and also within other membrane-bound compartments (e.g. nucleoli, nuclear speckles and Cajal bodies inside the nucleus) [8]. They are engaged in different cellular metabolism serving multiple functions, including cell signaling[9–12], biochemical catalysis[13–15], stress response/adaption[16–20], chromatin formation[21–23], muscle regeneration[24–26], and cellular noise buffering [27–29].

MOs display liquid-like properties, such as surface tension-driven spherical shape, fusion/fission capability, dripping/wetting behavior and rapid internal/external material exchange[8]. The dynamic, weak, multivalent molecular interactions of intrinsically disordered proteins (IDPs) and their interactions with binding partners (often RNA molecules) provide the driving force for the liquid-liquid phase separation (LLPS) of MOs [9,30–32]. The nature and biophysical properties of MOs are susceptible to multiple parameters, such as concentration, pH, temperature, ionic strength, binding partner, crowding agents and post-translational modifications (PTMs)[33]. Dysregulation of the phase behavior could generate aberrant phase transitions [34,35] and pathological structure formation[36,37], leading to neurodegenerative diseases (e.g. amyotrophic lateral sclerosis, frontotemporal dementia, Alzheimer disease, tauopathies, repeat expansion disorders), cancer, and infectious diseases[38]. As a facile and universal method of molecular organization in the cell environment, the high dynamics and multifaceted environmental responsiveness of MOs have aroused intense interests from cell biologists and material scientists alike. The intriguing features are attributed to the low specificity and large disorder of IDPs, which are distinctive from how classical proteinaceous systems are organized three dimensionally. Insights into MOs and their cognate molecular structures could help to not only elucidate underlying basis of design from nature but also generate inspirations for *de novo* design of novel biomaterials. MOs-inspired artificial systems therefore should contribute to the enhancement of knowledge about the cellular processes associated with phase changes that are integral to both physiology and pathology. In this review, we will first focus on materials designed by mimicking MOs at the molecular structure level (**Graphical Abstract** and **Figure 1**). Synthetic approaches by chemical and biological routes are introduced as methods for expanding this family of bio-inspired materials. The new material properties and MO-inspired functions will create exciting opportunities to expedite biochemical reactions, synthetic biology, drug discovery and drug delivery. These potential applications will be discussed in the third part of this review.

## 2. Design of membraneless organelles-inspired biomaterials

Intrinsically disordered proteins (IDPs) are the main constituents of MOs. Considered as the scaffolds of MOs, IDPs present some unique characteristics which contribute to the dynamic formation and disassembly of MOs and their enrichment in RNA. These characteristics include biased amino acids composition[39], weak and dynamic molecular interaction[40], multivalency[41] and RNA-interaction affinity[42]. An IDP is inter-spaced with low complexity (LC) domains/intrinsically disordered regions (IDRs), which display conformational disorder (lacking defined 3D structure) and low residue complexity. They are considered the modular molecular mediators of LLPS[36]. From naturally occurring IDPs, we can gain insights about the structural features of the protein design, as well as the functional motifs and the essential interactions found in LC domains, for the design of novel biomaterials. Representative examples of various biomaterials design, summarized in **Table 1**, will be further discussed in the following sections.

### 2.1. Engineered proteins

Proteins have been engineered based on IDPs to reveal the behavior of MOs and to construct bespoke materials with new functions. A general strategy incorporates repeats of LC domains/IDRs using synthetic biology to generate proteins capable of LLPS *in vivo* or *in vitro*. For example, Ddx4 and in particular, isolated N-terminal IDR of Ddx4 was used to reconstitute artificial MOs. Ddx4 as a primary constituent protein of nudge granules is responsible for forming membraneless organelles in living cells[43]. Electrostatic interactions were revealed as the leading driving forces for LLPS. Moreover, two structural characteristics, *i.e.*, presence of repeating blocks (8-10 residues) with alternating net charge and high enrichment of F/R-containing motifs (FG, GF, GR and RG) within positively charged blocks, were shown as the conserved features of LLPS-capable sequences. This feature was found to be prominent in 68 Ddx4 orthologs from 46 organisms[43]. The artificial MOs displayed LLPS, sensitive to temperature, ionic strength and arginine methylation. They were found to selectively concentrate single-stranded DNA while excluding double-stranded DNA. Thus, the artificial MOs imitated the natural MO's environmental responsiveness and preferential cargo recruitment.

Protein polymers inspired from Pro- and Gly-rich extracellular matrix IDPs (extracellular matrix proteins including resilin and elastin) with programmable and controllable lower/upper critical solution temperature (L/U-CST)[44] and hysteretic LLPS behavior[45] in physiological solutions were also investigated. Based on proteomic analysis of prototypical IDPs, P-$X_n$-G-containing motifs (where n varies from 0 to 4 and X is any amino acid) were proposed as the structural scaffold motifs to encode phase behavior owing to its large prevalence among IDPs. High enrichment of nonpolar residues and zwitterionic pairs were revealed for elastin and resilin, respectively. Resembling original IDPs in terms of amino acids nature and patterning, synthetic protein polymers could also exhibit similar type of phase behavior (L/U-CST). The hysteresis in phase behavior was investigated using polypeptides containing $(VAPVG)_n$ and derivatives as repeating units. The syntax (positioning of amino acids within a repeat) and the number of repeating units are two molecular determinants [45] of hysteresis. Subtle alteration (e.g., VGPVG and VAPVG, n=30) and sequence reversing (e.g., VPGVG and GVGPV, n=30) in amino acid syntax could drastically alter hysteresis. For $(VAPVG)_n$ system, hysteresis increases with the chain length, whilst a minimum n(*ca.* 40) is required to exhibit prominent hysteresis.

Another comprehensive study showed that six IDRs, including Lsm4, Tia1, Pub1, eIF4GII, hnRNPA1 and FUS, could self-assemble or co-assemble with RNA to generate reconstituted membraneless organelles[46]. Low salt concentration, presence of a crowding agent, and incorporation of RNA were identified as three promoters of LLPS of IDRs. Constructs made of IDRs fused with PTB (a RBP with multiple RNA binding domains), underwent maturation over time in the presence of RNA, eventually forming irreversible amyloid-like fibrillary structure. The liquid-to-solid transition and the resulting structures resembled a natural LLPS system from the condensation of hnRNP1 protein[36]. Apart from the demonstration of LLPS from full-length LAF-1 protein and isolated RGG domain (arginine/glycine-rich domain that is necessary and sufficient for LLPS[47]), the impact of polyadenylate RNA (poly-rA) of various lengths (from 15 to 3000 nt) on LAF-1 protein was also systematically investigated[48]. Short RNA molecules (poly-rA30 and poly-rA15) could weaken two-body interactions (that is, less-negative second virial coefficient) and increase internal mobility within the phase-separated droplets (that is, higher diffusion coefficient and lower viscosity). On the other hand, long RNA molecules (poly-rA3k) could strengthen effective two-body interactions and decrease the internal mobility within the droplets. Interestingly, the absolute concentration of protein in the droplet is revealed to be ultra-low, in the semi-dilute polymer solution regime with number density of $5 \times 10^{-5}$ molecules/nm$^3$, which is at least two orders of magnitude lower than the condensed phase of other IDP LLPS system[49]). The droplet appears to be relatively permeable, with an effective mesh size of 3-8 nm.

LC domains can be fused with other protein domains to impart new functions, leveraging on the phase separation behavior. Zhong et al[50] fused the TDP-43 LC domain with mussel foot protein 5 (Mfp5) to synthesize TDP43 LC-Mfp5 (TLC-M) protein, which underwent LLPS and subsequently solidified into amyloid-like fibers, thus giving rise to protein-based coating materials. Intriguingly, the coatings exhibited strong adhesion across a wide range of pH (3 to 11) and salt conditions (up to 1 M NaCl). Imposing new stimuli to trigger LLPS has been made possible by fusing LC domains with proteins responsive to the stimuli of interest. Using a similar approach, spatiotemporal control of LLPS in the living cell context could be achieved *via* an optogenetic method[51]. *Arabidopsis thaliana* Cry2, a blue-light-dependent self-associated-prone protein, was fused to the LC domains of FUS, Ddx4 and hnRNPA1 to construct light-sensitive phase-separated proteins. The triggered opto-droplets also underwent similar aging event (into irregular assemblies) over time as the native proteins[36,46].

The compositional control of MOs was systematically studied using a model system consisting of multivalent scaffolds and clients. Here, the multivalent scaffolds are formed by three independent interacting pairs constructed from at least 4 repetition of interaction motifs (including polySUMO/polySIM, polySH3/polyPRM and polyUCUCU/PTB). Clients are constructed from at most 3 repetition of interaction motifs (including SUMO/SIM, SH3/PRM and UCUCU/RRM with fluorescence protein tag)[52]. The multivalent scaffolds and clients represent the essential and dispensable components for the protein-based biocondensates formation, respectively. By comparing different pairs of scaffolds *in vitro*, Rosen et al showed that the free sites of scaffolds bind clients where the binding strength scales with client valency. The relative stoichiometries of pairs of scaffolds govern the client recruitment and phase behavior in a switch-like manner, namely, client recruitment changes steeply as stoichiometries/valences of scaffolds change, whilst client valency also impacts recruitment. A simple mass action mode was devised to explain the compositional regulation of MOs.

## 2.2. Oligopeptides self-assembly

In the intracellular LLPS milieu[53], the low complexity (LC) domain is dominated by polar, charged and aromatic amino acid residues. They allow weaker and more dynamic intermolecular interaction than hydrophobic residues. An atomic 'Velcro' effect is provided by low-complexity aromatic-rich kinked segments (LARKS)[39,40], which are sometimes called reversible amyloid cores (RACs) [54,55]. These peptide sequences provide the driving force for the condensation of some IDPs in cells, as uncovered by scanning the LC domain. Self-assembling peptides, usually 20 residues or shorter, can be designed based on LARKS. Variations include repetitions [44,45,56] and systematic mutations of the LC-derived sequences [57]. The assemblies of these peptides range from fibrils [40,54–59], phase separated droplets[44,45,54,56,57], to hydrogels[40,55,58,59], while some also display thermo-sensitivity [40,44,45,54,55] and thermo-reversibility[44,45,54]. The latter characteristics distinguish the MO-inspired peptide assemblies from the irreversible and much more stable peptide assemblies of highly ordered beta-sheets and cognate steric zipper originated from amyloid plaques[60–69].

Fused in Sarcoma (FUS) protein, an RBP implicated in the RNA-granules formation, was reported to harness multiple reversible amyloid cores (RACs) to mediate the dynamic assembly of the FUS LC domains[54]. RACs provide weaker interactions compared to the hydrophobicity-driven steric zipper structure in amyloid plaques [70]. Two tandem (S/G)Y(S/G) sequences, namely, $^{37}$SYSGYS$^{42}$ and $^{54}$SYSSYG$^{59}$, are capable of forming labile and thermosensitive fibrillary nanoscale architecture under physiological conditions. Consistently, FUS protein variants with deleted RAC motif exhibited largely diminished propensity for phase separation. It was further revealed that RAC-driven nanoscale fibrillary structure adopted an hydrophilic sheet interface, in contrast to the highly hydrophobic structure in the pathological amyloid-like fibrils.

Seven amyloid cores were investigated in TAR DNA-binding protein 43 (TDP-43)$^{39}$, one RNA binding protein engaged in the formation of both stress granules and amyloid. While six peptide motifs derived from the protein were pathogenic and aggregation-prone, $^{312}$NFGAFS$^{317}$ is an RAC motif capable of forming a dynamic, reversible, crystal-like nanostructure. It could be disassembled at 70 °C with 1% sodium dodecyl sulfate (SDS) treatment, while the pathogenic amyloid fibrils remained unchanged. Moreover, the stability of this RAC motif can be further modified by $^{315}$A mutation. Thus, $^{312}$NFGEFS$^{317}$ and $^{312}$NFGpTFS$^{317}$ exhibited highly stability, while lability was maintained for $^{312}$NFGTFS$^{317}$. Intriguingly, A-to-pT mutation enhanced the stability of the assemblies, owing to the increased area buried for the dry interface and enhanced molecular interdigitation, albeit the presence of electrostatic repulsion.

In another example, the RAC motifs of heterogeneous nuclear ribonucleoprotein (hnRNP) family were systematically investigated$^{55}$. RAC1 peptide ($^{209}$GFGGNDNFG$^{217}$) underwent thermosensitive and reversible fibrillary nanostructure formation under physiological-relevant conditions. As expected, protein mutants lacking the RAC motifs were undermined in their capacity to phase-separate. Two other peptides from TDP-43, MNFGAFSINP and EDLIIKGISV, were shown to form nanofibrils[58], while NFGAF and DLII were identified as the minimal fibrillogenic sequences[59]. The original (NFGAF and DLII) and the derived variations of DLII (YLII, KLII, NLII, and LIID) could form hydrogels containing fibrillary nanostructures and high water content (99.9%) when peptide concentrations in water were sufficiently high. They provide a new way to generate peptidic soft biomaterials.

While shorter peptides (<10 a.a.) are capable of self-assembling into nanofibrils, longer peptides (with at least 19[57], 20[56], 25[71], 36[72] a.a. lengths reported) alone derived from LC domains can undergo LLPS. This phenomenon was demonstrated by two LC domains, LC1 (231–308) and LC2 (317–407), from U1 small nuclear ribonucleoprotein 70 kDa (U1-70K, a protein that associates exclusively with tau in the cortex of Alzheimer's disease patients). Phase separated droplets composed of LC1 or LC2 were observed in the presence of PEG as a crowding agent [73]. Moreover, a nonadecapeptide ERERRRDRDRDRDRDREHK (RD19), derived from LC2 domain, phase-separated into spherical droplets at millimolar concentration in the presence of a crowding agent [57]. The D to A mutant (RA19) formed irregular "droplets" while R to A mutant (AD19) failed to display phase separation, where in RA19 and AD19, the acidic (negatively charged) and basic residues (positively charged) are substituted by alanine residues (neutral), respectively. The results indicated that the modulation of charge interaction is essential for LLPS. Histidine-rich squid beak proteins (HBPs), another type of IDP, undergoes LLPS under conditions with specific pH and ionic strength [71]. GHGXY (X is a hydrophobic residue) was identified as an important motif for LLPS from these proteins. The derived oligopeptides, GHGVYGHGVYGHGPYGHGPYGHGLY[71] and $(GHGLY)_n$ (where n is at least 4)[56], formed phase separated-droplets by self-coacervation without any crowding agent. As proposed in a molecular interaction model, histidine residues in the peptide sequence acted as a molecular 'switch', which became deprotonated during pH shift (from acidic to neutral conditions) and formed hydrogen bonding with tyrosine. The clustering of tyrosine then helped to stabilize the LLPS-driven microscale droplets. Note that at least 4 repeats of GHGXY are required for the peptide sequences to undergo LLPS, presumably owing to the necessity of intramolecular histidine-tyrosine interactions at high proximity for the initiation of LLPS[71]. Longer pharmaceutical peptides (including seven hybrid incretin peptides and one enfuvirtide with at least 36 a.a. lengths) were also revealed to undergo LLPS in the presence of PEG crowding agent at specific pH condition, whilst high peptide concentration and crowded solution environments are suggested as two favorable parameters for LLPS[72].

Keating et al reported complex coacervation from RNA and oligopeptide wherein electrostatic attraction was the main driving force[74]. Inspired by Kemptide (LRRASLG), a model synthetic substrate for protein kinase A (PKA) containing modular peptides (RRASL)$_n$ (n=2 or 3) was designed. PolyU and tRNA were shown to coacervate with cationic (RRASL)$_n$ oligopeptide to form model MOs. The charging state of serine and the cognate phase behavior could be reversibly controlled by kinase and phosphatase, thereby generating efficient enzymatic control in an artificial LLPS system.

## 2.3. Polyelectrolytes co-assembly

Natural MOs are largely comprised of polyelectrolytes varying in the nature and density of charge. In particular, the constituents of MOs include two pivotal categories of biological polyelectrolytes[75] – the polycationic IDPs and polyanionic RNAs [46,76]. Thus, the mimicry of MOs leads to the exploration of polyelectrolytes as building blocks of bio-inspired materials.

Charged amino acids are prevalent in the LC domain and contribute to the expanded conformation (distinctive from common compact conformations of proteins). Moreover, the conformation of IDP is largely dependent on the net charge of the protein and the ionic strength of the solution milieu[77]. Pappu et al studied the effect of the distribution of oppositely charged residues in a protein[78]. Considering a protein as a polyampholyte – a polymer harboring both positively and negatively charged repeating units, the nature of protein folding is first determined by both the fraction of charged residues. High fraction of charged residue generates strong polyampholyte, and *vice versa*. A weak polyampholyte adopts a globule-like conformation, while a strong polyampholyte's conformation is further dependent on the charged residues' distribution on the linear sequence. When oppositely charged residues are well-mixed, the protein has a random coil structure. In contrast, a protein with a more ordered segregation of oppositely charged residues adopt a in hairpin-like conformation. This work helps to elucidate the connection between sequence and the ensemble of polyelectrolytes, and guide the design of polyelectrolytes with tunable strength of molecular interaction and of different conformations.

The oppositely charged polyelectrolytes were systematically studied by Obermeyer et al[79]. Compared to a classical wide-type globular protein, a supercharged globular protein has a drastically enhanced capacity to complex with electrostatic complementary binding partners owing to surface charge modulation, especially at physiological pH and moderate ionic strength.

The authors expressed cationic supercharged green fluorescent protein (csGFP) and studied the phase behavior with anionic macromolecular binding partners, including both biological macromolecules (DNA and RNA) and synthetic macromolecules (poly(acrylic acid) and poly(styrenesulfonate)). Experimental conditions including mixing ratios, ionic strength and pH were systematically altered and examined, and the result indicates csGFP-synthetic macromolecules system phase separate at higher ionic strength conditions than biological ones. This work highlights supercharging as a new and facile route towards ionic complementary artificial LLPS system, while suggesting mixing ratios, ionic strength and pH as key parameters that modulate phase behavior.

Lindhoud et al reported a three-component coacervation system[80]. As a model LLPS system, oppositely charged polyelectrolytes pair, namely, poly(acrylic acid) (PAA, polyanionic, resembling RNA in terms of charging state) and poly(allylamine hydrochloride) (PAH, polycationic, resembling IDP in terms of charging state), was utilized to construct polyelectrolyte complexes (PECs). Compared with protein-based artificial MOs, this is a much simpler system and requires merely commonly available polymers. The partitioning behavior of lysozyme and succinylated lysozyme, two structurally similar but oppositely charged proteins, was studied. The PEC composition was quantified by charge ratio $F^-$, namely, negative (PAA) concentration divided by the sum of negative (PAA) and positive (PAH) charge concentrations. The protein mixture could be selectively separated under specific PECs composition. Under optimal partitioning charge ratio $F_{opt}^-$ ($F_{opt}^-$ =0.65 and 0.55 for lysozyme and succinylated lysozyme, respectively), the partitioning coefficient could reach up to $10^4$, thus suggesting strong preferential partitioning behavior. Selective partitioning could be reversed by altering pH (within range of 4 to 10), thus suggesting their potential applications in protein extraction and controlled release. A multi-component coacervation system was reported by Keating et al[81], as a model system to recapitulate the ubiquitous intracellular sub-compartmentalization, namely, the formation of demixing bio-condensates phases [82,83]. Up to three polycations were mixed with up to three polyanions to form complex coacervate that harbored up to three coexisting subcompartments. Distributions of subcompartments of polyelectrolytes (polycations/polyanions) within coacervates were dependent on the relative (electrostatic) interactions among them, which in turn affected the partitioning of oligonucleotide/oligopeptide cargos within the coacervates. For instance, in the double coacervate system formed from 2xRRASL, Prot, polyU, Glu100 (RRASLRRASL peptide, protamine sulfate, poly(uridylic acid) and poly(L-glutamic acid), respectively), the cargo A15 RNA oligonucleotide was more enriched in the 2xRRASL/Prot/polyU phase (inner subcompartment) compared to Glu100 phase (outer subcompartment), presumably owing to attraction of polyU *via* Watson−Crick base pairing.

**2.4. Hybrid of proteins/peptides and polymers**

Hybrid biomaterials from proteins/peptides and polymers could exhibit phase separation under certain conditions. Based on the molecular drivers of phase separation, this category can be subdivided into two groups: 1) synthetic polymers containing LLPS-driving segments conjugated to peptides/proteins, and 2) synthetic polymers grafted with peptides that provide the favorable interaction for LLPS.

N-isopropylacrylamide (NIPAAM)-based synthetic polymers [84] and grafted copolymers[85] were reported to exhibit temperature-dependent phase separation behavior and leveraged as structural units for the construction of protein-polymer hybrid materials. Hoffman et al synthesized hybrid materials from cytochrome $b_5$ and oligo(NIPAAM) (molecular weight 1.9 kDa, roughly 17 repeats)[86]. Cysteine-free cytochrome $b_5$ protein was genetically engineered by incorporating the threonine to cysteine (T→C) mutation *via* a precise site-directed mutagenesis, followed by maleimide-terminated oligo(NIPAAM) conjugation. The engineered hybrid materials exhibited comparable lower critical solution temperature (LCST) phase behavior as poly(NIPAAM) ($M_n$=290 kDa, PDI=3.5). Likewise, 3-mercaptopropionic acid-terminated poly(NIPAAM)-immunoglobulin G (IgG) hybrid with LCST phase behavior and antigen ceomplexation was reported by Kikuchi et al[87], which has potential as novel drug carriers and bioreactors. Park et al reported poly(NIPAAM)-trypsin hybrid with thermo-modulated enzymatic activity[88]. Thermo-responsive polymer poly(NIPAAM-co-GEMA) was fabricated from the co-polymerization of poly(NIPAAM) and glucosyoxylethyl methacrylate (GEMA), which was initiated by 4,4-azobis(4-cyanovaleric acid) (ACV) and harbored one carboxyl group. The polymer was subsequently conjugated with primary amine groups of trypsin, resulting in a polymer-coated enzymatic system, which underwent LCST phase transition behavior and exhibited a peculiar temperature-enzymatic activity relationship, namely, a doublet optimal temperature. Barner-Kowollik et al reported *in situ* synthesis of poly(NIPAAM)-bovine serum albumin (BSA) conjugates *via* one-step reversible addition-fragmentation chain transfer (RAFT) polymerization, thereby eliminating major postpolymerization purification steps[89]. Moreover, the hybrid exhibited unimpaired structural integrity, intact conformation-related esterase activity, and poly(NIPAAM) molecular weight-dependent LCST phase behavior. In the aforementioned examples, phase separation occurred with the condensation of the hydrophobic NIPAAM groups, liquid-solid phase separation (LSPS, precipitation) rather than LLPS was noted.

The LLPS in MO formation was revealed to be driven and governed *via* weak multivalent interactions with high reversibility, namely, interaction motifs ('stickers') spaced by flexible linkers ('spacers')[9315290]. This structural feature could be found in a myriad of synthetic polymer-peptide hybrid phase separation systems. These hybrids however were prepared at a time when there was no intent to mimic MO systems. Sogah et al fabricated a multiblock copolymer containing of GAGA (Gly-Ala-Gly-Ala) oligopeptide segments on a flexible poly(ethylene oxide)(PEO). The copolymer self-assembled into nanoparticles (LSPS) and the mechanical properties could be readily modulated by tuning building blocks[91]. van Hest et al reported pH-dependent LCST phase separation system from triblock copolymers[92]. Methacrylate-functionalized VPGVG peptide (Val-Pro-Gly-Val-Gly, as 'A' block) and methacrylate-functionalized E (glutamic acid, as 'B' block) were copolymerized into ABA triblock copolymers *via* atom transfer radical polymerization (ATRP). The synthetic copolymers built from short peptide motifs exhibited both LCST phase behavior and pH responsiveness, which were derived from the VPGVG blocks and E blocks, respectively. Another ABA triblock polymer was prepared using 2-isocyanatoethyl methacrylate (IEM) as block 'A' and PEG as block 'B', while the VPGVG-containing peptides were conjugated to the IEM block[93]. The LCST behavior could be readily tuned by altering the degree of polymerization, polymer concentration and pH, and the results indicate similar phase behavior as the linear poly(VPGVG), thereby short building blocks were found to be able to capture the non-covalent interactions of a much larger biomacromolecule essential for the phase behavior. In these studies, however, it was not completely clear whether LLPS or LSPS was observed.

Miller et al reported doubly thermo-responsive hydrogel from poly(NIPAAM)-oligopeptide (FEFEFKFK) hybrid[94]. Hybrid hydrogel formation was triggered once a concentration threshold was surpassed, and the storage modulus could be facilely modulated by altering the concentration of peptide content. Moreover, a doubly thermo-responsiveness was revealed. Upon heating, the hybrid hydrogel underwent an initial LCST phase transition (at 40°C, comparable to poly(NIPAAM)) and subsequent gel-sol transition (at 80°C), concomitant with gel storage modulus alteration, thereby indicating potential in regenerative medical applications. The interplay between phase transition and gelation was also investigated[95]. Similar to poly(NIPAAM), an LSCT phase behavior was observed for the polymer conjugated with peptides. The peptide modification was accompanied by more favorable enthalpic interaction for phase transition and a significant increase of the storage modulus (G') of the hydrogel formed, thereby indicating additional hydrogen bonding interactions between peptides and polymers.

We previously reported exploiting amyloid-inspired self-assembling peptide to construct tunable polymeric hybrid hydrogel via click chemistry method[96]. Leveraging the similar expertise, we recently demonstrated a minimalist approach for the synthesis of artificial MOs from dextran-oligopeptide hybrid[97]. Specific short oligopeptides were previously revealed to form thermo-responsive reversible fibrils[54], thereby indicating minimalist motifs could provide molecular attractions and drive LLPS. By extracting the molecular signatures, namely,

$^{37}$SYSGYS$^{42}$, from FUS protein, we prepared mimetics using simple chemical conjugation of short oligopeptides (<10 residues) to a dextran backbone. The IDP-mimicking polymer-oligopeptide hybrids (IPHs) phase-separate into micron-sized condensates with liquid-like properties under physiological conditions — reminiscent of natural MOs, whilst the LLPS propensity is dependent of molecular structures, namely, molecular weight (MW) of dextran, degree of modification (DM) of oligopeptide, and tyrosine/arginine (Y/R) ratio of oligopeptide. Furthermore, the IPH droplets are capable of dynamic recruitment and release of biomolecules and compartmentalization of enhanced biochemical reactions. This model system will help elucidate the molecular interactions implicated in MO formation and provide inspirations for the bottom-up design of biomimetic materials.

## 3. Applications

### 3.1. Artificial model systems for investigating properties of membraneless organelles

LLPS biomaterials can be used to reconstruct MOs in-vitro, that may serve as a study platform to quantitatively investigate unknown aspects of MOs. Examples include single and multiple recombinant protein component systems, polyelectrolytes mixtures, and RNA/peptide coacervates. Numerous synthetic membraneless condensates have been synthesized that display reversible formation and dissolution upon changes of temperature[49], ionic strength[98], pH[99], crowding condition[100–102], UV light[103], enzymes[74,104] and chemical reactions[105]. They have also been essential in studies elucidating the structural features of IDPs, as well as the biochemical and material properties of MOs. As such, using Ddx4-inspired proteinaceous bodies it was demonstrated that the droplet interior exhibited an organic solvent-like nature, that could destabilize and melt double-stranded DNAs, while providing preferential partitioning and stabilization of single stranded DNAs and RNAs[106]. The droplet milieu was also found to display viscosity significantly higher than that of water [47,48] and full of permeable voids with characteristic mesh size to control entry of intracellular molecules[48].

The multicomponent nature of MOs was captured using simple in-vitro models (**Figure 2**). An artificial LLPS platform, consisting of multiple IDPs, demonstrated the possibility of IDP-droplets (scaffolds) to recruit other IDPs (clients) with further formation of subcompartments and formation of segregated multiphase droplets due to the fusion of distinct IDP-droplets[107]. Correspondingly, Lopez et al[49] used 4 engineered ELPs, with various size and sequence composition, to produce microdroplets with multiple layers, analogous to the core-shell organization of stress granules and the hierarchical arrangement of nucleolus, demonstrating that the immiscibility of different LLPS systems can be dictated by the amino-acid sequence, protein chain length and molecular weight. A mixture of polyelectrolytes were also shown to form two- and three-phase droplets provided that pairs of oppositely charged macromolecules had dissimilar density [108]. Nucleic acid incorporating model systems, in particular native RBPs[27], RBP-inspired synthetic peptides (RP3 and SR8)[109] and synthetic construct of Fm-ferritin incorporating RNA-binding domain[110], provided experimental evidence that RNAs help to drive LLPS, multilayered organization, affect morphology and the material state of MOs.

Reconstituted systems with advanced functional mimicking were also reported. RGG domains in LAF-1 protein were utilized as the building blocks for the synthetic membraneless organelles[111], displaying selective recruitment and the controlled release of cargo. Zhang et al reported the reconstitution of postsynaptic density (PSD) *via* the LLPS of major PSD scaffold proteins (PSD-95, Shank, GKAP, and Homer) [112]. Multivalent interaction networks were fabricated both in solution and on supported membrane bilayers and the reconstituted PSD condensates could achieve multifaceted functional mimicking, including selective recruitment of glutamate receptors and synaptic enzymes, promotion of actin polymerization and repelling of inhibitory postsynaptic protein, thus suggesting a promising neuronal-synapse-mimicking molecular platform.

Arosio et al [113] applied microfluidics to create water-in-oil emulsion to synthesize 30 μm-150 μm "artificial cells" which host droplets composed of DEAD-box ATPase Dhh1 and RNA. These droplets were meant to mimic the processing bodies of yeast and allowed the authors to study the dynamics of their LLPS in a cell-like environment. The timescale of LLPS was shown to decrease linearly with the increase of cell volume. In addition, gravity-induced coarsening of the droplets (from coalescence and Ostwald ripening) could be arrested by introducing cytoskeleton-mimicking polymeric hydrogel matrix inside the artificial cell.

As the field of LLPS in the biological context is growing, there is a mounting interest in developing reliable MO-reconstructions for testing hypotheses, elucidating the interplay between material states and functions, and discerning mechanisms of pathological pathways, which will be discussed in Section 3.3. Furthermore, the development of minimalistic biomimetics with a low level of complexity is expected to facilitate quantitative investigations for uncovering the key determinants of phase transition in a cellular context.

## 3.2. Nano-to-Micro biochemical reactors

MOs serve as compartments that enable the segregation of contents within the intracellular space to provide spatiotemporal regulation of various biochemical processes. MO-inspired biomaterials have a potential to reconstruct chemical functionalities of natural organelles and thus can serve as a new class of artificial nano- and microscale bioreactors with a range of biomimetic properties. The field has seen many proposed models of bioactive reactors that support various processes involving proteins, nucleic acids, inorganic molecules, and multiple component reactions (**Figure 3**), overall demonstrating LLPS as an ingenuine strategy to compartmentalize and enrich diverse components.

Model MOs constructed from a mixture of peptides and mononucleotides proposed by Mann et al was demonstrated as a bioreactor that sequesters heterocyclic molecules and inorganic nanoparticle catalysts to enable the corresponding reactions[99]. Additionally, the droplet interior was found to be favorable for the preconcentration of globular proteins[114] and enzymes, as well as the enhancement of glucose phosphorylation[99]. Droplets might also exhibit a lower degree of partitioning or complete exclusion of molecules depending on the interplay between hydrophobicity, solubility and electrostatic factors, in addition to the shape and size of molecules.

Poly-(diallyldimethylammonium) (PDDA)/PAA microdroplets were characterized to selectively exclude unfolded proteins, while providing the uptake and storage of proteins in folded state[115]. Here, decrease in surface charge density along with the bulky size of unfolded molecules contributed to reduced interaction of protein with the droplet and its selective exclusion. Adjustable and reversible sequestration of small proteins, such as lysozyme was achieved in PAA/ poly(*N,N* dimethylaminoethyl methacrylate) (PDMAEMA) model system[116] by fine tuning the electrostatic interactions between droplet MO-components and protein as a function of ionic strength. Kuffner et al proposed a flexible strategy for the selective compartmentalization of enzymes into membraneless compartments using LC domains. Adenylate kinase (as a model enzyme) conjugated to the LC sequence was shown to localize into the liquid condensates, achieving up to 140-folds increase in concentration. In addition, the lower polarity of the droplet interior resulted in the recruitment of substrate molecules, altogether providing up to 5-folds enhancement of enzymatic activity[117,118]. PDDA/ATP liquid condensate system was also successfully utilized as a hub for multi-step enzymatic reaction of actinorhodin polyketide synthase complex, providing selective retention and stabilization of the required enzymes with additional reactivity enhancement attributed to concentration of components and fluidity of the droplet interior that ensured dynamic transport of reactants[119]. Highlighting synthetic MOs as a concentration hub, these studies suggest the synergic effect of a droplet interior and the local increase in enzyme concentration be responsible for the acceleration of reactions.

Membrane-free microdroplets have been used to modulate biochemical reactions involving nucleic acid molecules. For example, carboxymethyl dextran sodium salt/polylysine (CM-Dex/PLys) based model MO has been shown to spatially accelerate RNA catalysis, attributed to the localized enrichment of RNAs and low polarity of the droplet interior[120]. Likewise, Poudyal et al [121] demonstrated the ability of liquid coacervates constructed from poly-diallyldimethylammonium chloride and oligoadenosine or oligoaspartic acid to readily uptake small molecules (ligands, cofactors) and RNAs, where localized concentration of nucleic acid and stimulating cofactors provided a favorable environment for template-directed polymerization of nucleotides, binding between RNA-aptamers and specific ligands, and RNA cleavage by ribozyme and deoxyribozyme. Ribozyme catalysis rate can be further stimulated by tuning the strength of interaction of RNA with polycations, using polyanions of different size, identity and quantity to influence the RNA folding state [122].

Besides simple biochemical reactions, membrane-free droplets can be employed for the organization of complex reactions involving myriads of biomolecules of various nature. Deng et al [123] attained a spatial organization of in-vitro transcription (IVT) in polyU/spermine-based artificial MOs, demonstrating the phase-separated droplet as a favorable environment for the efficient partitioning of dsDNAs and the enrichment of IVT-components, comprised of enzymes and transcription factors necessary for the reaction. Attempts have also been made to demonstrate membrane-free coacervates as a site for cell-free gene expression. CM-Dex/Plys coacervates were reported to retain multicomponent genetic machinery and host successive transcription, translation and protein folding within the droplet phase[124]. Although the gene expression took place at a more rapid rate inside the coacervates, this trend was not sustainable after 12h due to protein aggregation caused by the reduction in free volume and increase in intermolecular interaction. Later, some studies reported liquid condensates that displayed a selective retention of individual components from the reaction pool. As such, recombinant ELP with RNA-binding domain was demonstrated to form MOs capable of reversible and selective encapsulation of mRNA transcripts within the pool of other components of *in vitro* transcription and translation (IVTT) [125]. RBD provided selective binding to mRNA and its successive sequestration within the liquid droplet, while tRNA and ribosomes were excluded due to little or no interaction with the ELP-RBD. Combined with reversible phase transition, mRNAs were spatially separated from IVTT-complex, thus temporarily suppressing translation[125]. Conversely, artificial MOs formed by natural IDPs (FUS and Ewsr1) achieved spatial enrichment of all components of translation machinery and were used to facilitate the translation of non-canonical amino-acids in cells [126].

In the field of metabolic engineering, LLPS compartments have been engineered in bacteria and yeast cells to confer spatiotemporal organization and aid de novo synthesis of chemicals. Avalos and his colleagues[127] demonstrated a light-dependent LLPS system to effectively redirect the flux of deoxyviolacein pathway in budding yeast cells. Five different light-switchable compartmentalization systems for use in yeast were established by directly expressing IDP-fused optogenic clustering tools. In E. coli cells, synthetic IDP-based membraneless compartments were shown as an effective strategy to sequester enzymes and host cascade metabolic reactions[128]. Chilkoti et al. presented temperature-controlled intracellular droplet formation in E.coli with spatial phase separation memory and proposed strategies to control the local catalytic efficiency of various enzymes inside droplets[129], setting an exciting new direction of synthetic subcellular compartmentalization in metabolic engineering.

### 3.3. Experimental platforms for drug discovery

Growing evidences suggest that MOs are susceptible to pathological phase transition in the development of neurodegenerative diseases[130] and cancer[131]. Aberrant phase transition, typically from LLPS to solid-like assemblies, could be triggered by disease-associated mutations[132,133]) or aging [46]. The altered biophysical properties (e.g. molecular diffusivity, shape, stiffness, inability to disassemble) affect normal cellular processes and are associated with the disease pathology [134]. Such discovery opens new possibilities for the disease treatment, in which MOs and their components arise as targets for therapeutic interventions. Biomimetic MO systems are well positioned to facilitate a better understanding of the mechanism underlying pathological phase transition and to provide an experimental evaluation platform for new therapeutics. Current findings suggest at least three potential directions for therapeutic interventions: tuning phase separation propensity, modulating material properties of coacervates, and regulating fiber nucleation using intrinsic and extrinsic factors [130].

So far, model biopolymer systems reconstructed from native IDPs substantially contributed to the molecular level understanding of phase separation errors. UBQLN2 biomimetic demonstrated the significance of single amino-acid modifications in the 'sticker' region in altering phase transition behavior and the biophysical properties of condensates[135], while FUS-based models revealed the effect of mutations in prion-like -domains[136] and such data may guide various drug design strategies, from replacing or degrading faulty proteins to inducing specific modifications on RNAs .

The mechanism of pathological aggregation of TDP43 was recapitulated using optogenic reconstructions of membraneless inclusions, associated with ALS and FTD pathogenesis[137]. This system consequently helped to demonstrate the ability of RNAs to prevent progressive aberrant and neurotoxic phase transitions of TDP-43 containing condensates. RNA was also reported in other studies as a potential modulator of biophysical state of MOs[27]. Perturbation of other cellular components were also revealed to affect phase separated structures. As such, ATP appears to exhibit concentration dependent control of LLPS of FUS, halting its transition to an aggregated state at higher concentration, while possibly promoting aggregation of MO components at low level[138]. Antimicrobial peptide LL-3 was presented as an external agent that impairs the kinetics and functionality of condensates and restricts enzymatic activity of sequestered enzymes[139], while work by Ghosh et al identified three classes of non-constituent macromolecular species that can promote or suppress phase separation and alter material properties, thus can potentially interfere with pathological transitions of LLPS structures [140]. In-vitro models using biomimetic MOs have also shown that their material properties and the propensity to shift from liquid-like state to solid aggregates are heavily influenced by the intracellular environment [101]. In relation to that, recent in-vivo studies determined mTORC1 as the main regulator of macromolecular crowding inside cells, suggesting that its malfunctioning evokes aberrant phase transition of MOs[141]. Along the same line, further studies elucidating effect of change in crowding condition on phase separation will help to reveal anti-crowding therapies for the treatment.

The role of MOs in cancer therapy have been revealed in model systems reconstituted from purified MED1, BRD4, SRSF2, HP1α, FIB1, and NPM1[142]. Antineoplastic drugs were shown to selectively partition into these condensates, where drugs seem to exhibit their therapeutic activity. Notably, changes in droplets properties directly affected drug concentration and activity. As such increase in MED-1 condensates volume was correlated with the reduction of the droplet concentration of tamoxifen and its diminished effectiveness, suggesting condensate-dependent drug resistance. These model MOs can thus help to optimize drug effectiveness and further investigations of condensates-drug interplay may direct advances in cancer therapeutics. In the near future, it is anticipated that model MO-systems, recapitulating essential structural and kinetic aspects of natural MOs will serve as an essential experimental platform to evaluate hypotheses regarding the progression of various debilitating diseases, and as assays for testing potential therapeutics.

### 3.4. Novel drug delivery vehicles

Formation of MOs represents a novel material organization paradigm with significant uniqueness, including liquid-like nature with high mobility, simplicity of hierarchical structure, high dynamics of molecular diffusion, reversibility, responsiveness to multifaceted stimuli (e.g. ionic strength, pH, temperature, post-translational modifications), thereby opening the prospects to engineer advanced drug delivery platforms over a wide spectrum of design specifications (**Figure 4**).

As such, drug vehicles can be made to respond to changes in temperature, pH, ligand binding and other external triggers allowing controlled assembly and cargo release. Thermally reversible phase behavior of elastin-like polypeptides (ELPs) has been exploited to construct in-situ forming injectable drug depot. It was achieved by virtue of the LCST phase behavior of ELPs around 37 °C, which is biodegradable and could undergo disaggregation and clearance over time. Biodistribution studies revealed a 25-fold longer half-life of ELPs (compared with a control group of ELP with similar molecular weight but no phase separation/aggregation at 37 °C ($T_t$>50 °C)), thus suggesting promise as depots for prolonged cargo release[143] As a cancer therapy system, thermo-sensitive ELP ($ELP_1$) was conjugated with therapeutic radionuclide iodine-131, followed by intratumoral administration into implanted tumor xenografts in nude mice[144]. $ELP_1$ exhibited significantly longer residence time and improved the antitumor efficacy compared with $ELP_2$ (negative control protein with similar molecular weight to $ELP_1$, but no phase separation and aggregation under experimental conditions).

Responsiveness to pH is an attractive mechanism for the controlled drug delivery that LLPS-prone biomaterials can provide. Miserez et al [145]constructed a glucose responsive insulin delivery system from a histidine-rich beak protein (DgHBP)-inspired DgHBP-2 peptide with insulin and glucose oxidase co-incorporated, where its pH-responsive domain enabled active transition between open and enclosed complexes. The system exhibited prominent phase separation within pH range of 7.4-9.5 and underwent rapid dissociation upon local acidification with concomitant release of insulin, owing to the presence of glucose via gluconic acid conversion triggered by glucose oxidase. Notably, DgHBP-2 coacervates displayed release kinetics dependent on glucose-level, namely, a pulsatile release profile of insulin when the glucose concentration was altered between normal (1 mg/mL) and hyperglycemic levels (4 mg/mL) every 1.5 h.

Incorporation of enzymatic trigger was demonstrated with RGG-based synthetic droplets [111]. This flexible platform incorporates cargo-binding motifs (coiled-coil pair SYNZIP1 and SYNZIP2) and multiple protease recognition sites (Glu-Asn-Leu-Tyr-Phe- Gln-Gly) to achieve protease-mediated controlled release of single and multiple protein cargos. The idea was demonstrated using green fluorescent protein (GFP, *ca*. 30 kDa) and red fluorescent protein (RFP, *ca*. 30 kDa). This study demonstrates the possibility of using IDP-mimetics for the design of enzyme-triggered drug delivery, while further modifications could be applied to expand their functionality as cargo delivery vehicles.

Dworak et al constructed a thermo-responsive poly(NIPAAM)-oligopeptide coacervate with site-specific enzymatic cleavage properties[146]. The dansyl-labeled GRKFG-conjugated poly(NIPAAM) nanoparticle was synthesized *via* ATRP, wherein the core and corona were composed of poly(NIPAAM) and GRKFG, respectively. In addition, enzymatically triggered peptide release was demonstrated *via* site-specific (at R or K residue of peptide) trypsin hydrolysis, with no appreciable change in sizes of nanoparticles.

Polyelectrolytes-based drug delivery system may offer some advantages, as a simple strategy to carry RNA and protein therapeutics. Moreover, their charged nature, which makes them extremely sensitive to changes in the charge state of its components, can be utilized to trigger drug release. One example of such system is complex coacervates formed by RNAs and cationic peptides with phosphorylatable serine sites [74]. This coacervate displayed reversible formation and dissolution in response to the phosphorylation enzymatic addition/removal of a phosphate group on the peptide. Oppositely charged polyelectrolyte systems also offer a versatile method for protein encapsulation. For example, poly(L-lysine) and poly(D/L-glutamic acid) were reported to form and encapsulate proteins through simple mixing in aqueous buffer condition [147].This system demonstrated efficient encapsulation of BSA without altering its secondary structure and a large range of tunability over encapsulation efficiency and release kinetics by changing the protein to polypeptide ratio and the molecular weight of the components. As an alternative LLPS strategy for protein encapsulation, recent study proposed polyionic coacervation tags, in the form of short amino-acid sequences [148] which could induce LLPS of globular proteins (green fluorescent protein and CAP). Design parameters, such as the number and location of charged residues in the tag were primarily based on the sequence characteristics of MO-forming proteins, demonstrating the utility of short coacervation tags for therapeutic delivery, protein stabilization and purification.

An important class of MOs known as ribonucleoprotein (RNP)-granules exhibits interesting features applicable for RNA-carrier design. These structures are known to be effective for RNA sequestration and potentially protective from nuclease degradation[149] Interactions between RNA and RNP-granules proteins (RBPs), likely driven by weak multivalent interactions[42], can easily be reversed, allowing instantaneous switch-like phase transition and rapid release of components. Classical RBPs are known to contain various RNA-binding domains, such as RNA-recognizing motif (RRM), zinc-fingers, RGG-domains [27,150,151], that drive their association with RNAs. In-vitro, RNA phase separation was achieved in the presence of both LC domains and RBDs[46] or in some cases with RBD alone [47]. Thus, the panel of RBP-subdomains can serve as a valuable tool for RNA-drug carrier design, contributing to the field of protein-based RNA delivery and newly emerged class of protein-based RNA carriers (including siRNA, miRNA and mRNA) [152]. Further investigation of key factors that determine the propensity for phase separation and RNA recruitment in MOs would provide guidelines for rational design of effective nucleic acid carriers.

Another inspiration comes from coexisting, heterogeneous liquid phases that underlie the sub-compartmentalized architecture of some MOs, which were reconstructed by synthetic components (Figure 2). This kind of system can advance the development of novel therapeutic platforms, spanning from stimuli-responsive sequential delivery and release of multiple drugs, mediating delivery of incompatible drugs and even as platform for in-situ drug synthesis[153]. While phase separating systems are highly stimuli-responsive and provide efficient means for drug/gene encapsulation, these vehicles are generally less stable compared to conventional drug delivery systems and might not be suitable for systemic delivery routes, thus finding strategies for effective stabilization will be the next step in expanding their utility as drug delivery vehicles. Incorporation of membrane enclosure is one of the directions, as recent studies reported ways to induce transition of membraneless coacervates into vesicles by mild changes in external conditions [154,155] and formation of stabilizing proteinaceous membrane via self-assembling protein-polymer conjugates[156].

## 4. Concluding remarks

To sum up, intense research interests, especially recently, have been focused on the fundamental research and bioinspired applications concerning membraneless organelles owing to not only the unique underlying biophysics (multivalent weak interaction and liquid-liquid phase separation) but also the featured material properties (liquidity, fluidity, reversibility and highly dynamic nature). Besides the wide implications in physiology and pathology in the biological context, the insight into structural basis and unique materials properties offered a myriad of inspirations for materials design, including reconstituted organelles, nano/micro biochemical reactors, novel delivery vehicles, and therapeutic interventions. We envisage the emerging liquid phase of membraneless organelles are engendering a new phase of not only fundamental and interdisciplinary research of biology but also rational design of a new generation of biomaterials.

## 5. Acknowledgement

The authors would like to thank the funding support from the Hong Kong Research Grants Council (GRF 16102520 and GRF 16103517). Jianhui Liu receives financial support from the Hong Kong PhD Fellowship Scheme. We thank BioRender.com for creating figures.

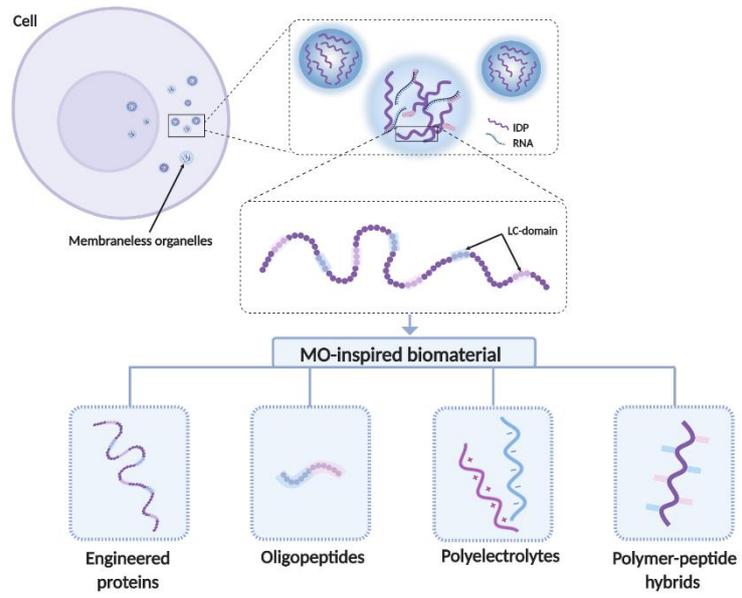

**Graphical abstract**

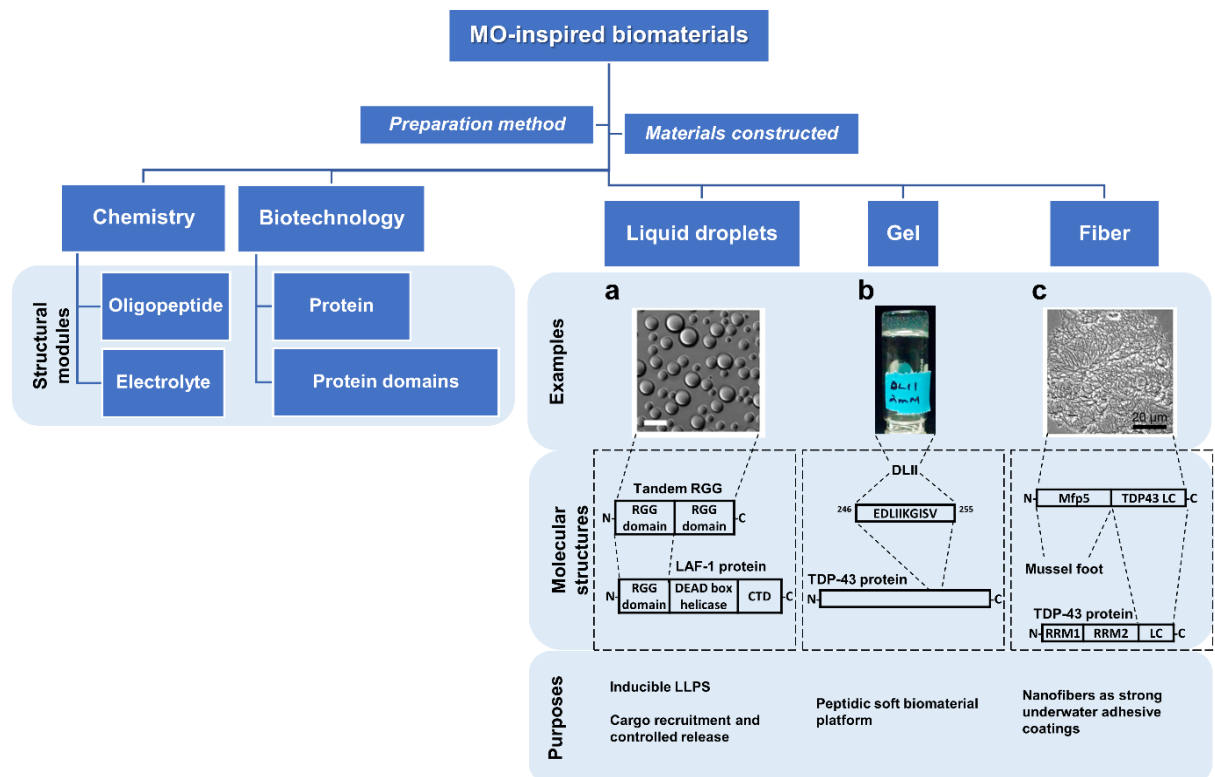

**Figure 1.** Multifaceted bio-inspired materials from membraneless organelles. Various routes could be leveraged for the construction of biomaterials, including both chemistry and biotechnology techniques. Biomaterials with distinct material states, including liquid droplet[111], gel[59] and fiber[50], could be fabricated for bespoke applications. Reprinted with permission from ref [59]. Copyright 2014 American Chemical Society.

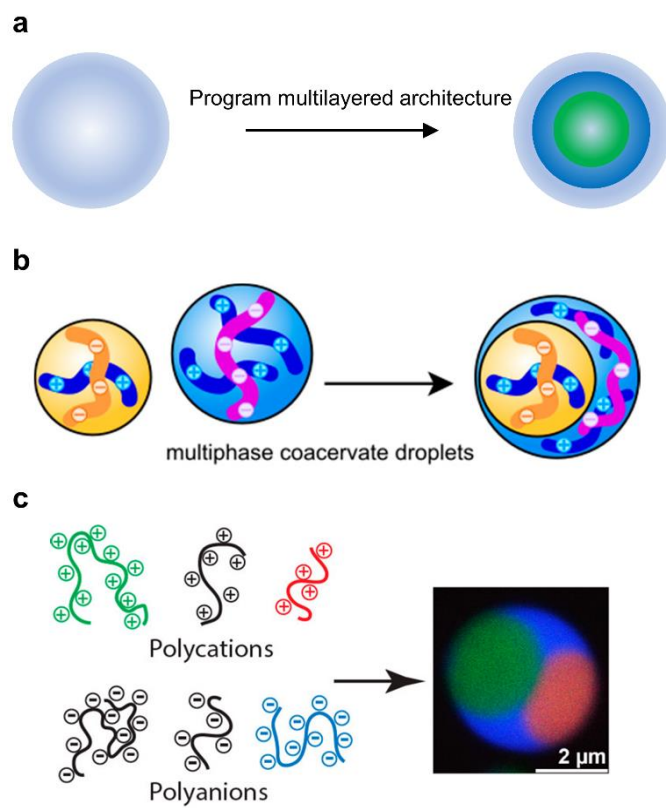

**Figure 2.** Strategies employed to reconstruct multiphase structure of MOs in-vitro: **a**, Temperature-triggered assembly of multilayered architecture from homogeneous droplets containing aqueous mixtures of ELPs encoded for coacervate self-assembly[49]. **b**, Formation of multiphase droplets in mixtures of complex coacervate droplets with different critical salt concentrations. Reprinted with permission from ref[108]. Copyright 2020 American Chemical Society. **c**, Sequential formation of multiphase droplet in the mixture of oppositely charged pairs of polyelectrolytes. Reprinted with permission from ref [81]. Copyright 2020 American Chemical Society.

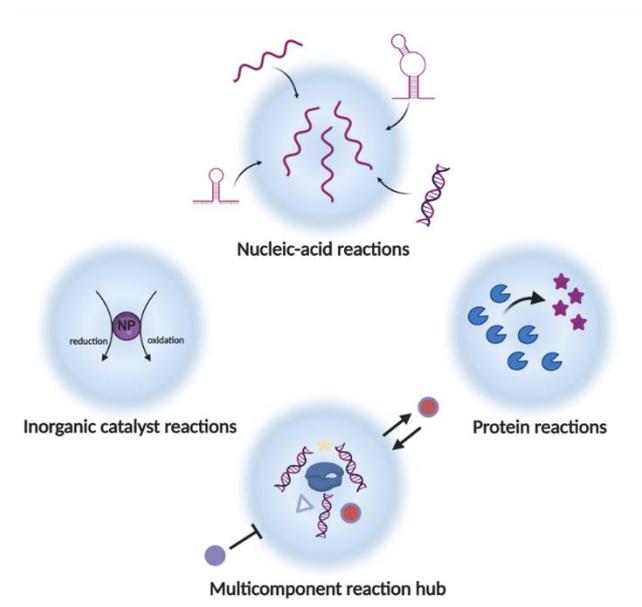

**Figure 3.** MO-inspired bioreactor systems.

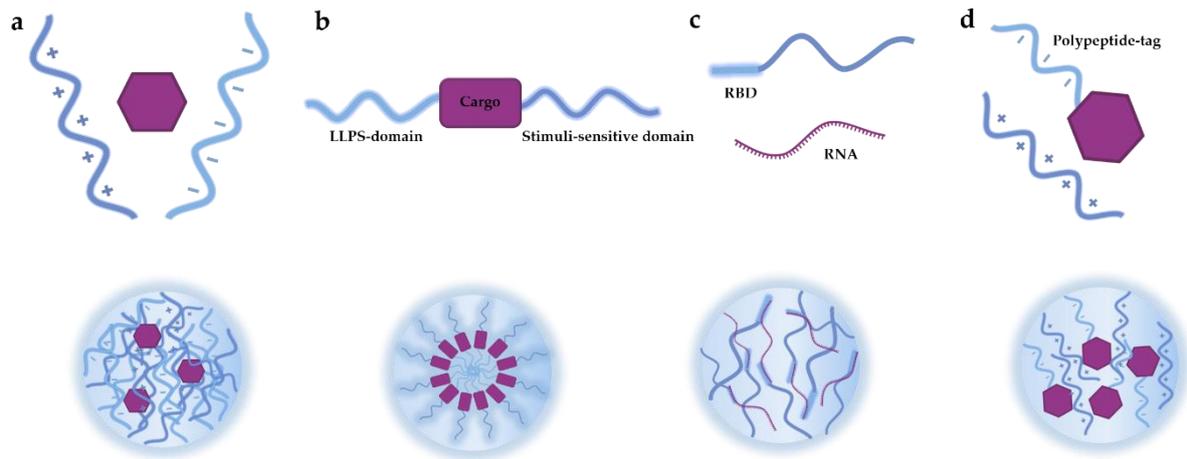

**Figure 4.** Strategies for LLPS drug delivery platform design: **a**, Polyelectrolytes-based assembly. **b**, Fusion of cargo molecules to LLPS and stimuli-sensitive domains. **c**, Incorporation of RNA-binding domain (RBD) for nucleic acid therapeutics encapsulation. **d**, Insertion of polypeptide tag to cargo molecules for polyanion-polycation based LLPS.

**Table 1**

Representative examples of biomaterials inspired from MOs.

| Composition | Origin of inspiration/commonality | Structural modules | Materials constructed | Ref |
|---|---|---|---|---|
| *Oligopeptide assembly* | | | | |
| Oligopeptide | FUS protein | Peptide motif: SYSGYS; SYSSYGQS; STGGYG | Reversible nanofibrils Labile hydrogels | 40 |
| Oligopeptide | FUS protein | Peptide motif: SYSGYS; SYSSYG | Reversible nanofibrils | 54 |
| Oligopeptide | TDP-43 protein | Peptide motif: Ac-MNFGAFSINP-NH$_2$; Ac-EDLIIKGISV-NH$_2$ | Nanofibrils | 58 |
| Oligopeptide | TDP-43 protein | Peptide motif: Ac-NFGAF-NH$_2$; Ac-DLII-NH$_2$ | Nanofibrils Hydrogels | 59 |
| Oligopeptide | TDP-43 protein | Peptide motif: NFGAFS | Reversible nanofibrils | 39 |
| Oligopeptide | hnRNP | Peptide motif: GFGGNDNFG; GFGNDGSNF; | Reversible nanofibrils Hydrogel formation for sequence: | 55 |
| | | YNDFGNY; GFGDGYNGYG; GYGGGYDNYGG; YNDFGNY; YGGDQNY; SDFQSN; SGYDYS; GYNNDN | GFGGNDNFG and GFGNDGSNF. | |
| Long pharmaceutical oligopeptide | Incretin peptide Enfuvirtide | Peptide motif including: HAibQGTFTSDYSKYLDERAAQDFVQWLLDGGPSSGAPPPSK-NH2; Ac-YTSLIHSLIEESQNQQEKNEQELLELDKWASLWNW F-NH$_2$ | Micron-sized peptide condensates | 72 |
| Oligopeptide Protein domains | U1-70K protein | LC domains: LC1 (231–308) and LC2 (317–407) Peptide motif: ERERRRDRDRDRDREHK | Micron-sized peptide droplets | 57 |
| Oligopeptide | Histidine-rich squid beak proteins | Peptide: | Micron-sized peptide droplets | 71 |
| | | DgHBP-pep (GHGVYGHGVYGHGPYGHGPYGHGLY) | | |
| Oligopeptide | Histidine-rich squid beak proteins | Motif: GHGLY Peptide: GHGLY(GAGFA)$_3$GHGLY; (GHGLY)$_4$; GHGLY(GHGLH)$_3$GHGLY; GHGLYGAGFAGHGLHGFAGHGLY | Micron-sized peptide droplets | 56 |
| Oligopeptide + RNA | Kemptide (LRRASLG) | Oligopeptide + RNA | Micron-sized droplets | 74 |
| *IDP-based* | | | | |
| Polypeptide protein polymer | Elastin/Resilin | Repetitive peptide motif: P–X$_n$–G (n varies from 0 to 4, and X is variable) | Protein particles | 44,45 |
| Protein /Protein domains | Ddx4 protein | LC domain | Protein droplets | 43 |
| Protein/Protein domains | Lsm4, Tia1, Pub1, eIF4GII, hnRNPA1 and FUS protein | LC domain | Protein droplets Maturated irregular aggregates/fibrils | 46 |
| Protein + RNA | LLPS of LAF-1 protein | Full-length protein RGG domain of protein | Micron-sized droplets | 48 |
| Protein with repetitive domains | SUMO, SIM, PRM, SH3, UCUCU and PTB motifs/domains | SUMO, SIM, PRM, SH3, UCUCU and PTB motifs/domains | Protein droplets | 52 |
| Reconstituted protein | Fus, Ddx4, hnRNP and Cry2 protein | LC domain (from Fus, Ddx4 and hnRNP) and light-activatable domain (Cry2) | Light-controllable protein droplets | 51 |
| Protein + RNA | DEAD-box ATPase Dhh1 | Protein + RNA | Cell-scale droplets hosting synthetic | 113 |

| | | | membraneless organelles | |
|---|---|---|---|---|
| Reconstituted protein | TDP-43 protein Mfp5 protein | TDP43 LC domain Mfp5 protein | Droplets (initial) and nanofibrils (final) Under water adhesive coatings | 50 |
| Protein with functional domains | LAF-1 | RGG domain + cargo-binding motif + protease cleavage site | Enzyme responsive droplets | 111 |
| RNA-binding protein | Elastin PGL-1 | ELP fused to RGG-domain from C-terminus of PGL-1 | Synthetic ribonucleoprotein-granules | 125 |
| Protein domains | DEAD-box proteins Dbp1 and Laf1 | LC-domains from N-terminus and C-terminus of Dbp1 and the N-terminus of LAF-1. | Droplet microreactors with selective recruitment of kinase enzyme | 117 |
| Protein /Protein domains | UBQLN2 | C-terminus of UBQLN2 (450-624) | Protein droplets | 135 |
| | | | | |
| *Polyelectrolyte-based* | | | | |
| Polyampholytic protein | Polyampholytic nature of IDPs | Strong polyampholyte system (Glu-Lys)25 with variable patterning | Macromolecules with variable radius of gyration ($R_g$) | 78 |
| Protein + synthetic polyelectrolyte | Complex coacervation in biology | Cationic supercharged green fluorescent protein (csGFP) Anionic synthetic macromolecules poly(acrylic acid) and poly(styrenesulfonate) | Micron-sized droplets Micron-sized assemblies | 79 |
| Ionic polypeptide tag | Complex coacervation in biology | GFP attached to [DEEEDD]$_n$-anionic tag + polycation poly(4-vinyl N-methyl pyridinium iodide | Globular protein coacervates | 148 |
| Synthetic polyelectrolytes | Complex coacervation in biology | poly(allylamine hydrochloride) / poly(acrylic acid) | Polyelectrolyte complex coacervates | 80 |
| | | poly(N,N dimethylaminoethyl methacrylate) /poly(acrylic acid) | Polyelectrolyte complex coacervates | 116 |
| Natural/synthetic polyelectrolytes | Complex coacervation in biology | Cationic polyelectrolyte: RRASLRRASL peptide, protamine sulfate, poly(L-lysine), poly(allylamine hydrochloride) Anionic polyelectrolyte: poly(acrylic acid), poly(L-glutamic acid), poly(L-aspartic acid) and poly(uridylic acid) | Micron-sized polyelectrolyte complexes harboring multi-subcompartments | 81 |
| | | Carboxymethyl dextran sodium salt /polylysine | Polyelectrolyte complexes | 120 |
| | | poly(diallyldimethylammonium)chloride, PDDA)/adenosine triphosphate (ATP) | Polyelectrolyte complexes | 114,119 |

| Synthetic peptide/RNA | RNP-granules | Two synthetic peptides (RP3 and SR8) + ssRNA | Droplets with RNA-mediated formation of dynamic substructures (vacuoles) | 109 |
|---|---|---|---|---|
| *Polymer hybrid with protein/peptide* | | | | |
| Modified protein | Multivalent molecular interactions in biology | Immunoglobulin G (IgG) modified with P(NIPAAm) | Phase separation | 87 |
| Modified protein | Multivalent molecular interactions in biology | Trypsin modified with P(NIPAAm) | Phase separation | 88 |
| Synthetic polypeptide polymer | Elastin | ABA triblock copolymers hosting VPGVG motifs | Phase separation | 92 |
| Synthetic polypeptide polymer | Elastin | ABA triblock copolymers grafted with VPGVG peptide side chain | Phase separation | 93 |

| Polymer-oligopeptide hybrid | FUS protein, 'sticker-and spacer' interaction mode | Short peptide motif: SYSGYS RGG | Liquid-liquid phase separation; micron-sized droplets; artificial membraneless organelles | 97 |

# 6. Reference

1. Wheeler, R. J. & Hyman, A. A. Controlling compartmentalization by non-membrane-bound organelles. *Philos. Trans. R. Soc. B Biol. Sci.* **373**, (2018).

2. Brangwynne, C. P. *et al.* Germline P Granules Are Liquid Droplets That Localize by Controlled Dissolution/Condensation. *Science* **324**, 1729–1732 (2009).

3. Uversky, V. N. Proteins without unique 3D structures: Biotechnological applications of intrinsically unstable/disordered proteins. *Biotechnol. J.* **10**, 356–366 (2015).

4. Shin, Y. & Brangwynne, C. P. Liquid phase condensation in cell physiology and disease. *Science* **357**, eaaf4382 (2017).

5. Alberti, S. Phase separation in biology. *Curr. Biol.* **27**, R1097–R1102 (2017).

6. Banani, S. F., Lee, H. O., Hyman, A. A. & Rosen, M. K. Biomolecular condensates: Organizers of cellular biochemistry. *Nat. Rev. Mol. Cell Biol.* **18**, 285–298 (2017).

7. Raposo, G. & Stoorvogel, W. Extracellular vesicles: Exosomes, microvesicles, and friends. *J. Cell Biol.* **200**, 373–383 (2013).

8. Hyman, A. A., Weber, C. A. & Jülicher, F. Liquid-Liquid Phase Separation in Biology. *Annu. Rev. Cell Dev. Biol.* **30**, 39–58 (2014).

9. Li, P. *et al.* Phase transitions in the assembly of multivalent signalling proteins. *Nature* **483**, 336–340 (2012).

10. Wright, P. E. & Dyson, H. J. Intrinsically disordered proteins in cellular signalling and regulation. *Nat. Rev. Mol. Cell Biol.* **16**, 18–29 (2015).

11. Chong, P. A. & Forman-Kay, J. D. Liquid–liquid phase separation in cellular signaling systems. *Curr. Opin. Struct. Biol.* **41**, 180–186 (2016).

12. Su, X. *et al.* Phase separation of signaling molecules promotes T cell receptor signal transduction. *Science* **352**, 595–599 (2016).

13. Sabari, B. R. *et al.* Coactivator condensation at super-enhancers links phase separation and gene control. *Science* **361**, eaar3958 (2018).

14. Boija, A. *et al.* Transcription Factors Activate Genes through the Phase-Separation Capacity of Their Activation Domains. *Cell* **175**, 1842-1855.e16 (2018).

15. Gallego, L. D. *et al.* Phase separation directs ubiquitination of gene-body nucleosomes. *Nature* **579**, 592–597 (2020).

16. Adriaens, C. *et al.* P53 induces formation of NEAT1 lncRNA-containing paraspeckles that modulate replication stress response and chemosensitivity. *Nat. Med.* **22**, 861–868 (2016).

17. Rabouille, C. & Alberti, S. Cell adaptation upon stress: the emerging role of membrane-less compartments. *Curr. Opin. Cell Biol.* **47**, 34–42 (2017).

18. Riback, J. A. *et al.* Stress-Triggered Phase Separation Is an Adaptive, Evolutionarily Tuned Response. *Cell* **168**, 1028–1040 (2017).

19. Franzmann, T. M. *et al.* Phase separation of a yeast prion protein promotes cellular fitness. *Science* **359**, eaao5654 (2018).

20. Franzmann, T. M. & Alberti, S. Protein phase separation as a stress survival strategy. *Cold Spring Harb. Perspect. Med.* **11**, a034058 (2019).

21. Strom, A. R. *et al.* Phase separation drives heterochromatin domain formation. *Nature* **547**, 241–245 (2017).

22. Rai, A. K., Chen, J. X., Selbach, M. & Pelkmans, L. Kinase-controlled phase transition

of membraneless organelles in mitosis. *Nature* **559**, 211–216 (2018).

23. Gibson, B. A. *et al.* Organization of Chromatin by Intrinsic and Regulated Phase Separation. *Cell* **179**, 470-484.e21 (2019).

24. Vogler, T. O. *et al.* TDP-43 and RNA form amyloid-like myo-granules in regenerating muscle. *Nature* **563**, 508–513 (2018).

25. McHugh, J. TDP-43 in the muscles: friend or foe? *Nat. Rev. Rheumatol.* **15**, 1–1 (2019).

26. Mori, F. *et al.* Phosphorylated TDP-43 aggregates in skeletal and cardiac muscle are a marker of myogenic degeneration in amyotrophic lateral sclerosis and various conditions. *Acta Neuropathol. Commun.* **7**, 1–12 (2019).

27. Maharana, S. *et al.* RNA buffers the phase separation behavior of prion-like RNA binding proteins. *Science* **360**, 918–921 (2018).

28. Riback, J. A. & Brangwynne, C. P. Can phase separation buffer cellular noise? *Science* **367**, 364–365 (2020).

29. Klosin, A. *et al.* Phase separation provides a mechanism to reduce noise in cells. *Science* **367**, 464–468 (2020).

30. Fasting, C. *et al.* Multivalency as a chemical organization and action principle. *Angew. Chemie Int. Ed.* **51**, 10472–10498 (2012).

31. Harmon, T. S., Holehouse, A. S., Rosen, M. K. & Pappu, R. V. Intrinsically disordered linkers determine the interplay between phase separation and gelation in multivalent proteins. *Elife* **6**, e30294 (2017).

32. Wang, J. *et al.* A Molecular Grammar Governing the Driving Forces for Phase Separation of Prion-like RNA Binding Proteins. *Cell* **174**, 688-699.e16 (2018).

33. Gomes, E. & Shorter, J. The molecular language of membraneless organelles. *J. Biol. Chem.* **294**, 7115–7127 (2019).

34. Alberti, S. & Hyman, A. A. Are aberrant phase transitions a driver of cellular aging? *BioEssays* **38**, 959–968 (2016).

35. Jain, A. & Vale, R. D. RNA phase transitions in repeat expansion disorders. *Nature* **546**, 243–247 (2017).

36. Molliex, A. *et al.* Phase Separation by Low Complexity Domains Promotes Stress Granule Assembly and Drives Pathological Fibrillization. *Cell* **163**, 123–133 (2015).

37. Peskett, T. R. *et al.* A Liquid to Solid Phase Transition Underlying Pathological Huntingtin Exon1 Aggregation. *Mol. Cell* **70**, 588-601.e6 (2018).

38. Alberti, S. & Dormann, D. Liquid–Liquid Phase Separation in Disease. *Annu. Rev. Genet.* **53**, 171–194 (2019).

39. Guenther, E. L. *et al.* Atomic structures of TDP-43 LCD segments and insights into reversible or pathogenic aggregation. *Nat. Struct. Mol. Biol.* **25**, 463–471 (2018).

40. Hughes, M. P. *et al.* Atomic structures of low-complexity protein segments reveal kinked β sheets that assemble networks. *Science* **359**, 698–701 (2018).

41. Mitrea, D. M. & Kriwacki, R. W. Phase separation in biology; Functional organization of a higher order Short linear motifs - The unexplored frontier of the eukaryotic proteome. *Cell Commun. Signal.* **14**, (2016).

42. Rhine, K., Vidaurre, V. & Myong, S. RNA Droplets. *Annu. Rev. Biophys.* **49**, 247–265 (2020).

43. Nott, T. J. *et al.* Phase Transition of a Disordered Nuage Protein Generates

Environmentally Responsive Membraneless Organelles. *Mol. Cell* **57**, 936–947 (2015).

44. Quiroz, F. G. & Chilkoti, A. Sequence heuristics to encode phase behaviour in intrinsically disordered protein polymers. *Nat. Mater.* **14**, 1164–1171 (2015).

45. Quiroz, F. G. *et al.* Intrinsically disordered proteins access a range of hysteretic phase separation behaviors. *Sci. Adv.* **5**, 5177–5195 (2019).

46. Lin, Y., Protter, D. S. W., Rosen, M. K. & Parker, R. Formation and Maturation of Phase-Separated Liquid Droplets by RNA-Binding Proteins. *Mol. Cell* **60**, 208–219 (2015).

47. Elbaum-Garfinkle, S. *et al.* The disordered P granule protein LAF-1 drives phase separation into droplets with tunable viscosity and dynamics. *Proc. Natl. Acad. Sci. U. S. A.* **112**, 7189–7194 (2015).

48. Wei, M. T. *et al.* Phase behaviour of disordered proteins underlying low density and high permeability of liquid organelles. *Nat. Chem.* **9**, 1118–1125 (2017).

49. Simon, J. R., Carroll, N. J., Rubinstein, M., Chilkoti, A. & López, G. P. Programming molecular self-assembly of intrinsically disordered proteins containing sequences of low complexity. *Nat. Chem.* **9**, 509–515 (2017).

50. Cui, M. *et al.* Exploiting mammalian low-complexity domains for liquid-liquid phase separation–driven underwater adhesive coatings. *Sci. Adv.* **5**, eaax3155 (2019).

51. Shin, Y. *et al.* Spatiotemporal Control of Intracellular Phase Transitions Using Light-Activated optoDroplets. *Cell* **168**, 159-171.e14 (2017).

52. Banani, S. F. *et al.* Compositional Control of Phase-Separated Cellular Bodies. *Cell* **166**, 651–663 (2016).

53. Muiznieks, L. D., Sharpe, S., Pomès, R. & Keeley, F. W. Role of Liquid–Liquid Phase

Separation in Assembly of Elastin and Other Extracellular Matrix Proteins. *J. Mol. Biol.* **430**, 4741–4753 (2018).

54. Luo, F. *et al.* Atomic structures of FUS LC domain segments reveal bases for reversible amyloid fibril formation. *Nat. Struct. Mol. Biol.* **25**, 341–346 (2018).

55. Gui, X. *et al.* Structural basis for reversible amyloids of hnRNPA1 elucidates their role in stress granule assembly. *Nat. Commun.* **10**, 1–12 (2019).

56. Gabryelczyk, B. *et al.* Hydrogen bond guidance and aromatic stacking drive liquid-liquid phase separation of intrinsically disordered histidine-rich peptides. *Nat. Commun.* **10**, 1–12 (2019).

57. Xue, S. *et al.* Low-complexity domain of U1-70K modulates phase separation and aggregation through distinctive basic-acidic motifs. *Sci. Adv.* **5**, eaax5349 (2019).

58. Saini, A. & Chauhan, V. S. Delineation of the core aggregation sequences of TDP-43 C-terminal fragment. *ChemBioChem* **12**, 2495–2501 (2011).

59. Saini, A. & Chauhan, V. S. Self-assembling properties of peptides derived from TDP-43 C-terminal fragment. *Langmuir* **30**, 3845–3856 (2014).

60. Reches, M. & Gazit, E. Casting metal nanowires within discrete self-assembled peptide nanotubes. *Science* **300**, 625–627 (2003).

61. Li, D. *et al.* Structure-based design of functional amyloid materials. *J. Am. Chem. Soc.* **136**, 18044–18051 (2014).

62. Dai, B. *et al.* Tunable assembly of amyloid-forming peptides into nanosheets as a retrovirus carrier. *Proc. Natl. Acad. Sci. U. S. A.* **112**, 2996–3001 (2015).

63. Wei, G. *et al.* Self-assembling peptide and protein amyloids: From structure to tailored

function in nanotechnology. *Chem. Soc. Rev.* **46**, 4661–4708 (2017).

64. Nelson, R. *et al.* Structure of the cross-β spine of amyloid-like fibrils. *Nature* **435**, 773–778 (2005).

65. Sawaya, M. R. *et al.* Atomic structures of amyloid cross-β spines reveal varied steric zippers. *Nature* **447**, 453–457 (2007).

66. Laganowsky, A. *et al.* Atomic view of a toxic amyloid small oligomer. *Science* **335**, 1228–1231 (2012).

67. Wu, H. & Fuxreiter, M. The Structure and Dynamics of Higher-Order Assemblies: Amyloids, Signalosomes, and Granules. *Cell* **165**, 1055–1066 (2016).

68. Boke, E. *et al.* Amyloid-like Self-Assembly of a Cellular Compartment. *Cell* **166**, 637–650 (2016).

69. Riek, R. & Eisenberg, D. S. The activities of amyloids from a structural perspective. *Nature* **539**, 227–235 (2016).

70. Murray, D. T. *et al.* Structure of FUS Protein Fibrils and Its Relevance to Self-Assembly and Phase Separation of Low-Complexity Domains. *Cell* **171**, 615-627.e16 (2017).

71. Tan, Y. *et al.* Infiltration of chitin by protein coacervates defines the squid beak mechanical gradient. *Nat. Chem. Biol.* **11**, 488–495 (2015).

72. Wang, Y., Lomakin, A., Kanai, S., Alex, R. & Benedek, G. B. Liquid-Liquid Phase Separation in Oligomeric Peptide Solutions. *Langmuir* **33**, 7715–7721 (2017).

73. Bai, B. *et al.* U1 small nuclear ribonucleoprotein complex and RNA splicing alterations in Alzheimer's disease. *Proc. Natl. Acad. Sci. U. S. A.* **110**, 16562–16567 (2013).

74. Aumiller, W. M. & Keating, C. D. Phosphorylation-mediated RNA/peptide complex

coacervation as a model for intracellular liquid organelles. *Nat. Chem.* **8**, 129–137 (2016).

75. Kayitmazer, A. B., Seeman, D., Minsky, B. B., Dubin, P. L. & Xu, Y. Protein-polyelectrolyte interactions. *Soft Matter* **9**, 2553–2583 (2013).

76. Weber, S. C. & Brangwynne, C. P. Getting RNA and protein in phase. *Cell* **149**, 1188–1191 (2012).

77. Müller-Späth, S. *et al.* Charge interactions can dominate the dimensions of intrinsically disordered proteins. *Proc. Natl. Acad. Sci. U. S. A.* **107**, 14609–14614 (2010).

78. Das, R. K. & Pappu, R. V. Conformations of intrinsically disordered proteins are influenced by linear sequence distributions of oppositely charged residues. *Proc. Natl. Acad. Sci. U. S. A.* **110**, 13392–13397 (2013).

79. Cummings, C. S. & Obermeyer, A. C. Phase Separation Behavior of Supercharged Proteins and Polyelectrolytes. *Biochemistry* **57**, 314–323 (2018).

80. Van Lente, J. J., Claessens, M. M. A. E. & Lindhoud, S. Charge-Based Separation of Proteins Using Polyelectrolyte Complexes as Models for Membraneless Organelles. *Biomacromolecules* **20**, 3696–3703 (2019).

81. Mountain, G. A. & Keating, C. D. Formation of Multi-Phase Complex Coacervates and Partitioning of Biomolecules Within Them. *Biomacromolecules* **21**, 630–640 (2020).

82. Feric, M. *et al.* Coexisting Liquid Phases Underlie Nucleolar Subcompartments. *Cell* **165**, 1686–1697 (2016).

83. Wheeler, J. R., Matheny, T., Jain, S., Abrisch, R. & Parker, R. Distinct stages in stress granule assembly and disassembly. *Elife* **5**, 18413 (2016).

84. Heskins, M. & Guillet, J. E. Solution Properties of Poly(N-isopropylacrylamide). *J. Macromol. Sci. Part A - Chem.* **2**, 1441–1455 (1968).

85. Chen, G. & Hoffman, A. S. Graft copolymers that exhibit temperature-induced phase transitions over a wide range of pH. *Nature* **373**, 49–52 (1995).

86. Chilkoti, A., Chen, G., Stayton, P. S. & Hoffman, A. S. Site-specific conjugation of a temperature-sensitive polymer to a genetically-engineered protein. *Bioconjug. Chem.* **5**, 504–507 (1994).

87. Takei, Y. G. *et al.* Temperature-Responsive Bioconjugates. 3. Antibody-Poly(N-isopropylacrylamide) Conjugates for Temperature-Modulated Precipitations and Affinity Bioseparations. *Bioconjug. Chem.* **5**, 577–582 (1994).

88. Lee, H. & Park, T. G. Conjugation of trypsin by temperature-sensitive polymers containing a carbohydrate moiety: Thermal modulation of enzyme activity. *Biotechnol. Prog.* **14**, 508–516 (1998).

89. Boyer, C. *et al.* Well-defined protein-polymer conjugates via in situ RAFT polymerization. *J. Am. Chem. Soc.* **129**, 7145–7154 (2007).

90. Martin, E. W. *et al.* Valence and patterning of aromatic residues determine the phase behavior of prion-like domains. *Science* **367**, 694–699 (2020).

91. Rathore, O. & Sogah, D. Y. Nanostructure formation through β-sheet self-assembly in silk-based materials. *Macromolecules* **34**, 1477–1486 (2001).

92. Ayres, L., Vos, M. R. J., Adams, P. J. H. M., Shklyarevskiy, I. O. & Van Hest, J. C. M. Elastin-based side-chain polymers synthesized by ATRP. *Macromolecules* **36**, 5967–5973 (2003).

93. Ayres, L., Koch, K., Adams, P. H. H. M. & Van Hest, J. C. M. Stimulus responsive behavior of elastin-based side chain polymers. *Macromolecules* **38**, 1699–1704 (2005).

94. Maslovskis, A., Tirelli, N., Saiani, A. & Miller, A. F. Peptide-PNIPAAm conjugate based hydrogels: Synthesis and characterisation. *Soft Matter* **7**, 6025–6033 (2011).

95. Maslovskis, A. *et al.* Self-assembling peptide/thermoresponsive polymer composite hydrogels: Effect of peptide-polymer interactions on hydrogel properties. *Langmuir* **30**, 10471–10480 (2014).

96. Liu, J., Ni, R. & Chau, Y. A self-assembled peptidic nanomillipede to fabricate a tuneable hybrid hydrogel. *Chem. Commun.* **55**, 7093–7096 (2019).

97. Liu, J., Zhorabek, F., Dai, X., Huang, J. & Chau, Y. Minimalist design of polymer-oligopeptide hybrid as intrinsically disordered protein-mimicking scaffold for artificial membraneless organelle. *arXiv Prepr. arXiv2103.15541* (2021).

98. Aumiller, W. M., Pir Cakmak, F., Davis, B. W. & Keating, C. D. RNA-Based Coacervates as a Model for Membraneless Organelles: Formation, Properties, and Interfacial Liposome Assembly. *Langmuir* **32**, 10042–10053 (2016).

99. Koga, S., Williams, D. S., Perriman, A. W. & Mann, S. Peptide-nucleotide microdroplets as a step towards a membrane-free protocell model. *Nat. Chem.* **3**, 720–724 (2011).

100. Marianelli, A. M., Miller, B. M. & Keating, C. D. Impact of macromolecular crowding on RNA/spermine complex coacervation and oligonucleotide compartmentalization. *Soft Matter* **14**, 368–378 (2018).

101. Kaur, T. *et al.* Molecular crowding tunes material states of ribonucleoprotein condensates. *Biomolecules* **9**, 71 (2019).

102. Lemetti, L. *et al.* Molecular crowding facilitates assembly of spidroin-like proteins through phase separation. *Eur. Polym. J.* **112**, 539–546 (2019).

103. Martin, N. *et al.* Photoswitchable Phase Separation and Oligonucleotide Trafficking in DNA Coacervate Microdroplets. *Angew. Chemie Int. Ed.* **58**, 14594–14598 (2019).

104. Saito, M. *et al.* Acetylation of intrinsically disordered regions regulates phase separation. *Nat. Chem. Biol.* **15**, 51–61 (2019).

105. Donau, C. *et al.* Active coacervate droplets as a model for membraneless organelles and protocells. *Nat. Commun.* **11**, 1–10 (2020).

106. Nott, T. J., Craggs, T. D. & Baldwin, A. J. Membraneless organelles can melt nucleic acid duplexes and act as biomolecular filters. *Nat. Chem.* **8**, 569–575 (2016).

107. Jo, Y. & Jung, Y. Interplay between intrinsically disordered proteins inside membraneless protein liquid droplets. *Chem. Sci.* **11**, 1269–1275 (2020).

108. Lu, T. & Spruijt, E. Multiphase Complex Coacervate Droplets. *J. Am. Chem. Soc.* **142**, 2905–2914 (2020).

109. Banerjee, P. R., Milin, A. N., Moosa, M. M., Onuchic, P. L. & Deniz, A. A. Reentrant Phase Transition Drives Dynamic Substructure Formation in Ribonucleoprotein Droplets. *Angew. Chemie Int. Ed.* **56**, 11354–11359 (2017).

110. Garcia-Jove Navarro, M. *et al.* RNA is a critical element for the sizing and the composition of phase-separated RNA–protein condensates. *Nat. Commun.* **10**, 3230 (2019).

111. Schuster, B. S. *et al.* Controllable protein phase separation and modular recruitment to form responsive membraneless organelles. *Nat. Commun.* **9**, 1–12 (2018).

112. Zeng, M. *et al.* Reconstituted Postsynaptic Density as a Molecular Platform for Understanding Synapse Formation and Plasticity. *Cell* **174**, 1172-1187.e16 (2018).

113. Linsenmeier, M. *et al.* Dynamics of Synthetic Membraneless Organelles in Microfluidic Droplets. *Angew. Chemie Int. Ed.* **58**, 14489–14494 (2019).

114. Williams, D. S. *et al.* Polymer/nucleotide droplets as bio-inspired functional micro-compartments. *Soft Matter* **8**, 6004–6014 (2012).

115. Martin, N., Li, M. & Mann, S. Selective Uptake and Refolding of Globular Proteins in Coacervate Microdroplets. *Langmuir* **32**, 5881–5889 (2016).

116. Lindhoud, S. & Claessens, M. M. A. E. Accumulation of small protein molecules in a macroscopic complex coacervate. *Soft Matter* **12**, 408–413 (2015).

117. Küffner, A. M. *et al.* Acceleration of an Enzymatic Reaction in Liquid Phase Separated Compartments Based on Intrinsically Disordered Protein Domains. *ChemSystemsChem* **2**, (2020).

118. Faltova, L., Küffner, A. M., Hondele, M., Weis, K. & Arosio, P. Multifunctional Protein Materials and Microreactors using Low Complexity Domains as Molecular Adhesives. *ACS Nano* **12**, 9991–9999 (2018).

119. Crosby, J. *et al.* Stabilization and enhanced reactivity of actinorhodin polyketide synthase minimal complex in polymer–nucleotide coacervate droplets. *Chem. Commun.* **48**, 11832–11834 (2012).

120. Drobot, B. *et al.* Compartmentalised RNA catalysis in membrane-free coacervate protocells. *Nat. Commun.* **9**, (2018).

121. Poudyal, R. R. *et al.* Template-directed RNA polymerization and enhanced ribozyme

catalysis inside membraneless compartments formed by coacervates. *Nat. Commun.* **10**, 1–13 (2019).

122. Poudyal, R. R., Keating, C. D. & Bevilacqua, P. C. Polyanion-Assisted Ribozyme Catalysis Inside Complex Coacervates. *ACS Chem. Biol.* **14**, 1243–1248 (2019).

123. Deng, N. N. & Huck, W. T. S. Microfluidic Formation of Monodisperse Coacervate Organelles in Liposomes. *Angew. Chemie - Int. Ed.* **56**, 9736–9740 (2017).

124. Tang, T.-Y. D., Van Swaay, D., DeMello, A., Ross Anderson, J. L. & Mann, S. In vitro gene expression within membrane-free coacervate protocells. *Chem. Commun.* **51**, 11429–11432 (2015).

125. Simon, J. R., Eghtesadi, S. A., Dzuricky, M., You, L. & Chilkoti, A. Engineered Ribonucleoprotein Granules Inhibit Translation in Protocells. *Mol. Cell* **75**, 66-75.e5 (2019).

126. Reinkemeier, C. D., Girona, G. E. & Lemke, E. A. Designer membraneless organelles enable codon reassignment of selected mRNAs in eukaryotes. *Science* **363**, (2019).

127. Zhao, E. M. *et al.* Light-based control of metabolic flux through assembly of synthetic organelles. *Nat. Chem. Biol.* **15**, 589–597 (2019).

128. Wei, S. P. *et al.* Formation and functionalization of membraneless compartments in Escherichia coli. *Nat. Chem. Biol.* 1–6 (2020). doi:10.1038/s41589-020-0579-9

129. Dzuricky, M., Rogers, B. A., Shahid, A., Cremer, P. S. & Chilkoti, A. De novo engineering of intracellular condensates using artificial disordered proteins. *Nat. Chem.* **12**, 814–825 (2020).

130. Elbaum-Garfinkle, S. Matter over mind: Liquid phase separation and neurodegeneration.

*J. Biol. Chem.* **294**, 7160–7168 (2019).

131. Bouchard, J. J. *et al.* Cancer Mutations of the Tumor Suppressor SPOP Disrupt the Formation of Active, Phase-Separated Compartments. *Mol. Cell* **72**, 19-36.e8 (2018).

132. Qamar, S. *et al.* FUS Phase Separation Is Modulated by a Molecular Chaperone and Methylation of Arginine Cation-π Interactions. *Cell* **173**, 720-734.e15 (2018).

133. Patel, A. *et al.* A Liquid-to-Solid Phase Transition of the ALS Protein FUS Accelerated by Disease Mutation. *Cell* **162**, 1066–1077 (2015).

134. Spannl, S., Tereshchenko, M., Mastromarco, G. J., Ihn, S. J. & Lee, H. O. Biomolecular condensates in neurodegeneration and cancer. *Traffic* **20**, 890–911 (2019).

135. Yang, Y., Jones, H. B., Dao, T. P. & Castañeda, C. A. Single Amino Acid Substitutions in Stickers, but Not Spacers, Substantially Alter UBQLN2 Phase Transitions and Dense Phase Material Properties. *J. Phys. Chem. B* **123**, 3618–3629 (2019).

136. McAlary, L., Plotkin, S. S., Yerbury, J. J. & Cashman, N. R. Prion-Like Propagation of Protein Misfolding and Aggregation in Amyotrophic Lateral Sclerosis. *Front. Mol. Neurosci.* **12**, 1–21 (2019).

137. Mann, J. R. *et al.* RNA Binding Antagonizes Neurotoxic Phase Transitions of TDP-43. *Neuron* **102**, 321-338.e8 (2019).

138. Kang, J., Lim, L. & Song, J. ATP enhances at low concentrations but dissolves at high concentrations liquid-liquid phase separation (LLPS) of ALS/FTD-causing FUS. *Biochem. Biophys. Res. Commun.* **504**, 545–551 (2018).

139. Oliva, R., Mukherjee, S. K., Fetahaj, Z., Möbitz, S. & Winter, R. Perturbation of liquid droplets of P-granule protein LAF-1 by the antimicrobial peptide LL-III. *Chem.*

*Commun.* **56**, 11577–11580 (2020).

140. Ghosh, A., Mazarakos, K. & Zhou, H.-X. X. Three archetypical classes of macromolecular regulators of protein liquid-liquid phase separation. *Proc. Natl. Acad. Sci.* **116**, 19474–19483 (2019).

141. Delarue, M. *et al.* mTORC1 Controls Phase Separation and the Biophysical Properties of the Cytoplasm by Tuning Crowding. *Cell* **174**, 338-349.e20 (2018).

142. Klein, I. A. *et al.* Partitioning of cancer therapeutics in nuclear condensates. *Science* **368**, 1386–1392 (2020).

143. Betre, H. *et al.* A thermally responsive biopolymer for intra-articular drug delivery. *J. Control. Release* **115**, 175–182 (2006).

144. Liu, W. *et al.* Injectable intratumoral depot of thermally responsive polypeptide-radionuclide conjugates delays tumor progression in a mouse model. *J. Control. Release* **144**, 2–9 (2010).

145. Lim, Z. W., Ping, Y. & Miserez, A. Glucose-Responsive Peptide Coacervates with High Encapsulation Efficiency for Controlled Release of Insulin. *Bioconjug. Chem.* **29**, 2176–2180 (2018).

146. Trzebicka, B. *et al.* Thermosensitive PNIPAM-peptide conjugate-Synthesis and aggregation. *Eur. Polym. J.* **49**, 499–509 (2013).

147. Black, K. A. *et al.* Protein encapsulation via polypeptide complex coacervation. *ACS Macro Lett.* **3**, 1088–1091 (2014).

148. Kapelner, R. A. & Obermeyer, A. C. Ionic polypeptide tags for protein phase separation. *Chem. Sci.* **10**, 2700–2707 (2019).

149. Lathe, R. & Darlix, J. L. Prion protein PrP nucleic acid binding and mobilization implicates retroelements as the replicative component of transmissible spongiform encephalopathy. *Arch. Virol.* **165**, 535–556 (2020).

150. Kim, Y. & Myong, S. RNA Remodeling Activity of DEAD Box Proteins Tuned by Protein Concentration, RNA Length, and ATP. *Mol. Cell* **63**, 865–876 (2016).

151. Schwartz, J. C., Wang, X., Podell, E. R. & Cech, T. R. RNA Seeds Higher-Order Assembly of FUS Protein. *Cell Rep.* **5**, 918–925 (2013).

152. Wang, W. & Tai, W. RNA binding protein as monodisperse carriers for siRNA delivery. *Med. Drug Discov.* **3**, 100011 (2019).

153. Elani, Y., Law, R. V. & Ces, O. Vesicle-based artificial cells as chemical microreactors with spatially segregated reaction pathways. *Nat. Commun.* **5**, 1–5 (2014).

154. Garenne, D. *et al.* Sequestration of Proteins by Fatty Acid Coacervates for Their Encapsulation within Vesicles. *Angew. Chemie - Int. Ed.* **55**, 13475–13479 (2016).

155. Jang, Y. *et al.* Understanding the Coacervate-to-Vesicle Transition of Globular Fusion Proteins to Engineer Protein Vesicle Size and Membrane Heterogeneity. *Biomacromolecules* **20**, 3494–3503 (2019).

156. Li, J., Liu, X., Abdelmohsen, L. K. E. A., Williams, D. S. & Huang, X. Spatial Organization in Proteinaceous Membrane-Stabilized Coacervate Protocells. *Small* **15**, 1–9 (2019).